\documentclass[aps,superscriptaddress,showpacs,nofootinbib,eqsecnum]{revtex4}

\usepackage{amssymb}  
\usepackage{amsmath}  
\usepackage{natbib}
\usepackage{epsfig}  
\usepackage{graphicx}  
\usepackage{dcolumn}
\usepackage{bm} 
\usepackage[english]{babel}  
\usepackage[latin1]{inputenc}

\begin{document}

\title{Numerical Solutions for non-Markovian Stochastic Equations of Motion}

\author{R. L. S.  Farias} \email{ricardo@dft.if.uerj.br}
\affiliation{Departamento de F\'{\i}sica Te\'orica, Universidade do Estado do
  Rio de Janeiro, 20550-013 Rio de Janeiro, RJ, Brazil}

\author{Rudnei O.  Ramos} \email{rudnei@uerj.br} \affiliation{Departamento de
  F\'{\i}sica Te\'orica, Universidade do Estado do Rio de Janeiro, 20550-013
  Rio de Janeiro, RJ, Brazil}

\author{L. A. da Silva} \email{las.leandro@gmail.com}
\affiliation{Departamento de F\'{\i}sica Te\'orica, Universidade do Estado do
  Rio de Janeiro, 20550-013 Rio de Janeiro, RJ, Brazil}

\begin{abstract}

The reliability and precision of numerically solving stochastic 
non-Markovian equations by standard numerical codes, more specifically, 
with the fourth-order Runge-Kutta routine for solving differential 
equations, is gauged by comparing the results obtained from analytical 
solutions for the equations. The results for different prescriptions 
for transforming the non-Markovian equations in a system of Markovian 
ones are compared so to check the reliability of the numerical method.

\bigskip
\bigskip

{\bf Published in:} Comp. Phys. Comm. {\bf 180}, 574 (2009)

\end{abstract}


\maketitle

\section{Introduction}
\label{sec1}

Stochastic equations of motion and theirs generalizations are extensively used
in different contexts, e.g. in classical statistical mechanics to study
systems with dissipation and noise, to determine how order parameters
equilibrate, in critical phenomena dynamics, among many other problems
\cite{reviews}.

The simplest case of a stochastic equation of motion is the phenomenological
Langevin equation describing e.g. the Brownian motion. It consists of
local (Markovian) dissipation and noise terms, which are related to each other
through the
classical Fluctuation-Dissipation theorem. Memory effects are then neglected.
However, in real systems scattering events responsible to dissipation and
fluctuations proceed through finite time intervals, which, consequently,
result in finite memory effects for these quantities.  The typical equations
of motion describing real physical systems are then expected to be nonlocal
(i.e., non-Markovian) ones with memory effects.  An equation of this type is
given by a generalized Langevin equation (GLE) of motion of the form (for a
general
reviews, see e.g.  Ref.~\cite{weiss}),

\begin{eqnarray}
  \ddot{\phi}(t)   +  \int_{0}^{t}   dt' K (t-t')\dot{\phi}(t')+
  V'(\phi) = \xi (t)\;,
  \label{genlangevin}
\end{eqnarray}
where $\phi$ is a variable of the system (for example the coordinate of a
particle) in interaction with a thermal bath (the dot means derivative
with respect to time),
$V(\phi)$ is a potential term
that is a function of $\phi$ (with prime denoting derivative with respect to
$\phi$), $K(t-t')$ is the dissipation kernel and the noise term $\xi(t)$ is a
Gaussian fluctuation with zero mean but colored, i.e., with two-point
correlation satisfying the generalized classical Fluctuation-Dissipation
relation at temperature $T$ (throughout this work we consider the Boltzmann
constant equal to one),

\begin{eqnarray}
  \left\langle \xi (t)\xi (t' )\right\rangle=
  TK (t-t' ).
  \label{genxixi}
\end{eqnarray}

Models with equations of the form of Eq. (\ref{genlangevin}) are also known
as Caldeira-Leggett type of models \cite{caldeira}.
Applications making use of equations of the form of Eq.~(\ref{genlangevin})
must be able to properly deal with the memory kernel. In some restrict cases,
like in the most common forms of non-Markovian kernels used in the literature,
e.g.  when the dissipation kernel $K(t-t')$, describes an Ornstein-Uhlenbeck
(OU) process~\cite{hanggi} or in the case of the Exponential Damped Harmonic
(EDH) kernel~\cite{alemaes1}, it is possible to reduce the non-Markovian
equation into a set of Markovian ones with white noise properties. In other
generic cases, however, this may not always be possible (see
Ref.~\cite{Luczka} for a recent review on the different colored noises and
associated equations used in the literature).  In these and any other cases
dealing for example, with nonlinear equations, we must resort to numerical
methods.  Though there are some specific numerical methods that may be
applicable for general cases \cite{bao1}, we still would like to be able to
solve equations like Eq.~(\ref{genlangevin}) through standard methods, which
are less numerically expensive than other alternatives.  Analytically, if the
equation is linear, then it is, in principle, possible to solve equations like
Eq.~(\ref{genlangevin}) through a Laplace transform, since the dissipation
integral term in Eq.~(\ref{genlangevin}) is just in the form of a convolution.
Even so, in these cases we can only look at averaged (over the noise)
quantities.  In addition, if it is nonlinear, i.e., when the potential term
$V(\phi)$ is a polynomial form of order larger than two in $\phi$, then we
must resort to numerical methods in order to solve the differential stochastic
equation of motion.  In the cases it can be reduced to a set of local
differential equations with white noise, like in the OU and EDH cases
mentioned above, the most natural way for solving the system of differential
equations would be e.g., through a standard Runge-Kutta method, which is both
easy to implement and usually produces results with good accuracy.  However,
since the standard Runge-Kutta method is basically a deterministic algorithm,
when applied to a stochastic differential equation with white noise, it
becomes not well defined, since the white noise term has infinite variance
and, therefore, cannot be generated. In order to deal with this problem,
stochastic Runge-Kutta routines have been proposed~\cite{rebeca}. An issue
related to applying such techniques to solve systems of differential equations
is that not all equations may be stochastic, but only a sub-set of them. In
cases like these, those algorithms may not be suitable or appropriate.

In this work our objective is to investigate both analytically and numerically
the solutions of non-Markovian linear equations of the form of
Eq.~(\ref{genlangevin}) and then compare their results.  In doing so, we are
able
to gauge the reliability and precision of numerically solving the stochastic
non-Markovian equations by standard numerical codes, e.g., Runge-Kutta codes
for solving differential equations.  {}For this, we use the most common forms
for the dissipation kernel $K(t-t')$ given by the OU and EDH forms.

The paper is organized as follows. In Section 2, we briefly describe
the solution of the linear GLE through Laplace transform.
In Section 3, we show the transformation of the GLEs of motion
with OU and EDH kernels in systems of Markovian time differential
equations with white noise. In Section 4, we show the comparison
of the results obtained for the analytical solutions for the linear
equations with those obtained numerically within our prescription
to make then local, from a standard fourth-order Runge-Kutta
code. {}Finally, in Section 5 we present our conclusions and final
comments about the precision of numerically solving the equations
with standard numerical methods.


\section{The generalized Langevin equation: linear regime}
\label{sec2}

Since non-Markovian equations like Eq.~(\ref{genlangevin}) have nonlocal
kernel terms in the form of a convolution, they become suitable to be solved
by Laplace transform. If we write the potential $V(\phi)$ in the form

\begin{equation}
V(\phi) = \frac{m^2}{2} \phi^2 + V_I(\phi)\;,
  \label{pot}
\end{equation}
where $m^2$ is a parameter of the potential and we have separated the
interaction term (non-quadratic) $V_I(\phi)$ from the quadratic one.  By
neglecting
interaction terms in Eq.~(\ref{pot}), the GLE Eq.~(\ref{genlangevin}) takes the
linear form,

\begin{eqnarray}
\ddot{\phi}(t)   +  m^2 \phi(t) + \int_{0}^{t}   dt' K (t-t')\dot{\phi}(t')
= \xi (t)\;.
\label{eom_linear}
\end{eqnarray}

By making use of the Laplace transform for $\phi(t)$,

\begin{eqnarray}
\mathcal{L}\{\phi(t)\} = \tilde{\phi}(s)
\equiv \int^{\infty}_{0} dt \, \exp(-st) \phi(t)\; ,
\end{eqnarray}
and from the convolution theorem applied to the non-Markovian dissipation term
in Eq.~(\ref{eom_linear}), we can easily obtain that the solution for the
linear GLE can be written in the Laplace transform
form as

\begin{equation}
\tilde{\phi}(s) = \frac{\dot{\phi}(0) +
\left[ s + \tilde{K}(s)\right]\phi(0)}{s^2
+ m^2 + s \tilde{K}(s)} + \frac{\tilde{\xi}(s)}{s^2
+ m^2 + s \tilde{K}(s)}\; ,
\label{phis}
\end{equation}
where $\tilde{K}(s)$ and $\tilde{\xi}(s)$ are the Laplace transforms of the
dissipation kernel $K(t-t')$ and the noise $\xi(t)$, respectively.

The solution for $\phi(t)$ is obtained from the inverse transform of
Eq.~(\ref{phis}),

\begin{equation}
\phi(t) =  \mathcal{L}^{-1}\{ \tilde{\phi}(s)\}
= \varphi(t) + \int^{t}_{0} dt' g(t-t') \xi(t') \; ,
\label{sol}
\end{equation}

\noindent
where

\begin{equation}
\varphi(t) = \mathcal{L}^{-1} \left\{ \frac{\dot{\phi}(0)
+ \left[ s + \tilde{K}(s)\right]\phi(0)}{s^2 + m^2
+ s \tilde{K}(s)} \right\} \; ,
\label{varphi}
\end{equation}
and

\begin{equation}
g(t-t') = \mathcal{L}^{-1} \left\{  \frac{1}{s^2 + m^2
+ s \tilde{K}(s)} \right\} \;.
\label{gtt}
\end{equation}

The explicit solution for $\phi(t)$ is difficult to give analytically because
of the noise term on the right hand side of Eq.~(\ref{sol}), but since the
noise is Gaussian, $\langle\xi\rangle=0$, we obtain that its average is simply
given by

\begin{equation}
\langle \phi(t) \rangle = \varphi(t)\;,
\label{avphi}
\end{equation}

\noindent
and once the kernel $K(t-t')$ is given, it is easily computed through
Eq.~(\ref{varphi}), either numerically or algebraically. We here have used the
{\it MAPLE} software to numerically evaluate for $\varphi(t)$. It should be
noted that in the OU and EDH cases the explicit forms for the solutions can be
obtained by {\it MAPLE}, but they are too complicated and long solutions, so
we refrain ourselves here to write them down explicitly.

It is also convenient to calculate $\langle\phi^2(t)\rangle$. Remembering that
the non-Markovian noise $\xi(t)$ satisfies Eq.~(\ref{genxixi}),
$\langle\xi(t)\xi(t')\rangle=T K(t-t')$, we then also obtain that

\begin{equation}
\langle \phi^2(t) \rangle = \varphi^2(t) +
T \int^{t}_{0} dt'' g(t-t'') \int^{t}_{0} dt' g(t-t')K(t'-t'') \; .
\label{avphi2}
\end{equation}


\section{The OU and EDH kernel cases}
\label{sec3}

Let us now describe the two cases of dissipation/noise kernels we are
interested in studying here, as mentioned in the introduction, the OU and EDH
cases, which are also the most common forms of non-Markovian kernels used in
the literature.  We will describe the equations with these two types of
kernels separately.

\subsection{The GLE with OU kernel}

Many studies considering the influence of Gaussian colored noise on nonlinear
physical systems are usually made considering the OU noise, with two-point
correlation satisfying

\begin{eqnarray}
\langle \xi_{OU}(t) \xi_{OU}(t^{\prime }) \rangle &=&
T K_{OU}(t-t^{\prime})\;,
\label{xitOU}
\end{eqnarray}
with \cite{hanggi}

\begin{equation}
K_{OU}(t-t^{\prime })=Q\gamma e^{-\gamma (t-t^{\prime })}\;,
\label{kernel_OU}
\end{equation}
where $\gamma$ gives the inverse of the time scale for the kernel memory and
$Q$ is the overall magnitude of the dissipation, which also gives the
magnitude of the dissipation in the local (Markovian) limit \cite{hanggi},

\begin{equation}
Q = \int_0^{\infty} dt' K(t-t')\;.
\label{eta}
\end{equation}

It can be easily shown that the OU noise can be generated by the stationary
part of the solution of the following differential equation:

\begin{equation}
\dot{\xi}_{OU}(t) = -\gamma\left[\xi_{OU}(t) - \sqrt{2TQ}
\zeta\right]\;,  \label{EDO_OU}
\end{equation}

\noindent
where $\zeta$ in Eq.~(\ref{EDO_OU}) is a white Gaussian noise satisfying

\begin{eqnarray}
\langle \zeta(t) \rangle &=&0\;,  \nonumber \\
\langle \zeta(t) \zeta(t^{\prime }) \rangle &=& \delta(t-t^{\prime})\;,
\label{zeta}
\end{eqnarray}

If we now define the new variable $W_{OU}(t)$, given by

\begin{equation}
W_{OU}\left( t\right) = -\int_{0}^{t}dt^{\prime }
K_{OU}\left(t-t^{\prime }\right) \dot{\phi}(t^{\prime })\;,
\label{WOU}
\end{equation}
it can be shown that $W_{OU}$ satisfies the equation of motion

\begin{equation}
\dot{W}_{OU}\left( t\right) = - \gamma W_{OU}\left(t\right) -
K_{OU}\left( 0\right) \dot{\phi}\left( t^{\prime }\right)\;.
\label{dotW_OU}
\end{equation}

Using Eqs. (\ref{EDO_OU}) and (\ref{dotW_OU}), we can transform the
integro-differential Eq.~(\ref{genlangevin}) with OU kernel into a
fourth-order dynamical system of local first-order differential equations
given by

\begin{eqnarray}
\dot{\phi} &=& y\;,  \nonumber \\
\dot{y} &=& - V^{\prime}(\phi) + \xi_{OU}
+ W_{OU} \;, \nonumber \\
\dot{W}_{OU} &=& -\gamma
W_{OU}- K_{OU}\left( 0\right) \dot{\phi} \;,
\nonumber \\
\dot{\xi}_{OU} &=& -\gamma\left[\xi_{OU} - \sqrt{2TQ}\,\,
\zeta\right]\;.
\label{setOU}
\end{eqnarray}


\subsection{The GLE with EDH kernel}

Another case of non-Markovian equation of interest in the literature is the
one with a EDH kernel, whose noise term, $\xi_{H}(t)$, satisfies

\begin{eqnarray}
  \langle \xi_{H}(t) \rangle &=&0\;,  \nonumber \\
  \langle \xi_{H}(t) \xi_{H}(t^{\prime }) \rangle &=& T K_{H}(t-t^{\prime
  })\;,  \label{xitH}
\end{eqnarray}
with kernel $K_{H}\left( t-t^{\prime }\right) $ given by

\begin{equation}
  K_H(t-t^{\prime })=e^{-\gamma(t-t^{\prime })}\frac{Q\Omega _{0}^{2}
  }{2 \gamma }\left\{ \cos [\Omega _{1}(t-t^{\prime })]+\frac{\gamma }{\Omega
      _{1}}\sin [\Omega _{1}(t-t^{\prime })]\right\} \;,
  \label{kernel_H}
\end{equation}
where $Q$ and $\gamma$ have the same meaning as in the OU case and
$\Omega_1^2~=~\Omega_0^2~-~\gamma^2~>~0$.  It can be easily shown that the
noise $\xi_H$ can be generated by the following differential
equation~\cite{alemaes1}:

\begin{equation}
  \ddot{\xi}_H(t)+ 2 \gamma \dot{\xi}_H(t)+\Omega _{0}^{2}\xi_H (t)=
\Omega _{0}^{2}  \sqrt{2TQ}\,\zeta(t)\;,
  \label{xieom}
\end{equation}
where $\zeta(t)$ is a white Gaussian noise with the same properties as given
in the OU noise case, expressed by Eq.~(\ref{zeta}).

Similarly as in the OU case, by defining a new variable $W_H$ analogous to
Eq.~(\ref{WOU}) and after analogous algebra leading to the system of equations
(\ref{setOU}), we obtain that the GLE with EDH
kernel can be written in terms of a sixth-order dynamical system of local
first-order differential equations given by

\begin{eqnarray}
  \dot{\phi}&=&y\;,  \nonumber \\
  \dot{y}&=&-V^{\prime }(\phi )+W_{H} + \xi_H \;,  \nonumber \\
  \dot{W}_{H}&=&u - 2 \gamma W_{H} - K_H(0)y\;,  \nonumber \\
  \dot{u}&=&-\Omega _{0}^{2} W_{H} +\dot{K}_{H}(0)y - 2 \gamma K_{H}(0)y\;,
  \nonumber \\
  \dot{\xi}_{H}&=&z\;,  \nonumber \\
  \dot{z}&=&-2\gamma z-\Omega _{0}^{2}\xi _{H} + \Omega _{0}^{2}\sqrt{2TQ}\,
  \zeta\;,
\label{setEDH}
\end{eqnarray}
where we have also defined a new function $u(t)$ as

\begin{equation}
u(t) = \int_{0}^t dt^{\prime }\left[ \frac{dK_H(t-t^{\prime })}{dt'} -
2 \gamma K_H(t-t^{\prime }) \right] \frac{d\phi(t^{\prime })}{dt'} \;.
\label{ut}
\end{equation}


\subsection{Alternative Prescription}

A different prescription than the one leading to the system of differential
equations (\ref{setOU}) and (\ref{setEDH}) is, instead of defining the
function $W(t)$ like in Eq.~(\ref{WOU}) (and similarly with the EDH kernel
case), would be to define it as

\begin{equation}
W\left( t\right) = -\int_{0}^{t}dt^{\prime }
K \left(t-t^{\prime }\right) \dot{\phi}(t^{\prime }) + \xi(t)\;,
\label{Wxi}
\end{equation}
whose only difference with the previous prescription is the addition of the
noise term to the equation. This prescription is used with some frequency in
the literature, e.g., like in \cite{bao2}.

In terms of Eq.~(\ref{Wxi}), the system of differential equations,
Eq.~(\ref{setOU}), for the OU case then changes to\footnote{Note that in
the OU case, with
the choice
Eq. (\ref{Wxi}), the noise $\xi_{OU}$ decouples from the system and
only the white noise contribution remains.}

\begin{eqnarray}
\dot{\phi} &=& y\;,  \nonumber \\
\dot{y} &=& - V^{\prime}(\phi)
+ W_{OU} \;, \nonumber \\
\dot{W}_{OU} &=& \gamma\sqrt{2TQ}\,\,\zeta -\gamma
W_{OU} - K_{OU}\left( 0\right)y\;,
\label{setOU2}
\end{eqnarray}
while Eq.~(\ref{setEDH}) for the EDH case changes to

\begin{eqnarray}
  \dot{\phi}&=&y\;,  \nonumber \\
  \dot{y}&=&-V^{\prime }(\phi )+W_{H} \;,  \nonumber \\
  \dot{W}_{H}&=&u - 2 \gamma (W_{H} - \xi_H) - K_H(0)y +z\;,  \nonumber \\
  \dot{u}&=&-\Omega _{0}^{2} (W_{H} - \xi_H) +\dot{K}_{H}(0)y -
2 \gamma K_{H}(0)y\;,
  \nonumber \\
  \dot{\xi}_{H}&=&z\;,  \nonumber \\
  \dot{z}&=&-2\gamma z-\Omega _{0}^{2}\xi _{H} + \Omega _{0}^{2}\sqrt{2TQ}\,
  \zeta\;.
\label{setEDH2}
\end{eqnarray}
The two systems of differential equations, Eqs. (\ref{setOU2}) and (\ref{setEDH2}),
are equivalent to the previous two, Eqs. (\ref{setOU}) and (\ref{setEDH2}), and they
produce identical results. The main difference between the two
prescriptions being the fact that the latter requires some extra
care in its numerical implementation. Since the function $W$, as
defined by Eq. (\ref{Wxi}), involves the noise function $\xi(t)$, its initial condition
must be carefully set in terms of stationary solution of the
noise differential equations, Eqs. (\ref{EDO_OU}) and (\ref{xieom}), for the OU and EDH
cases, respectively, otherwise the resulting dynamics between the
two different prescriptions for deriving the two sets of differential
equations will lead to different results.


\section{Comparing analytical and numerical solutions}
\label{sec4}

Now we show our analytical results for $\langle\phi(t)\rangle$ and
$\langle\phi^2(t)\rangle$ obtained using Laplace transformation and compare
with our numerical results obtained by solving the systems of equations
derived before from the two prescriptions used in the previous section, i.e.,
Eqs.  (\ref{setOU}) and (\ref{setEDH}) and Eqs.  (\ref{setOU2}) and
(\ref{setEDH2}). As commented at the end of the last section, the two prescriptions
lead to identical results. For the results shown below we used the
first prescription, leading to the systems of differential equations,
Eqs. (\ref{setOU}) and (\ref{setEDH}).

\begin{figure}[htb]
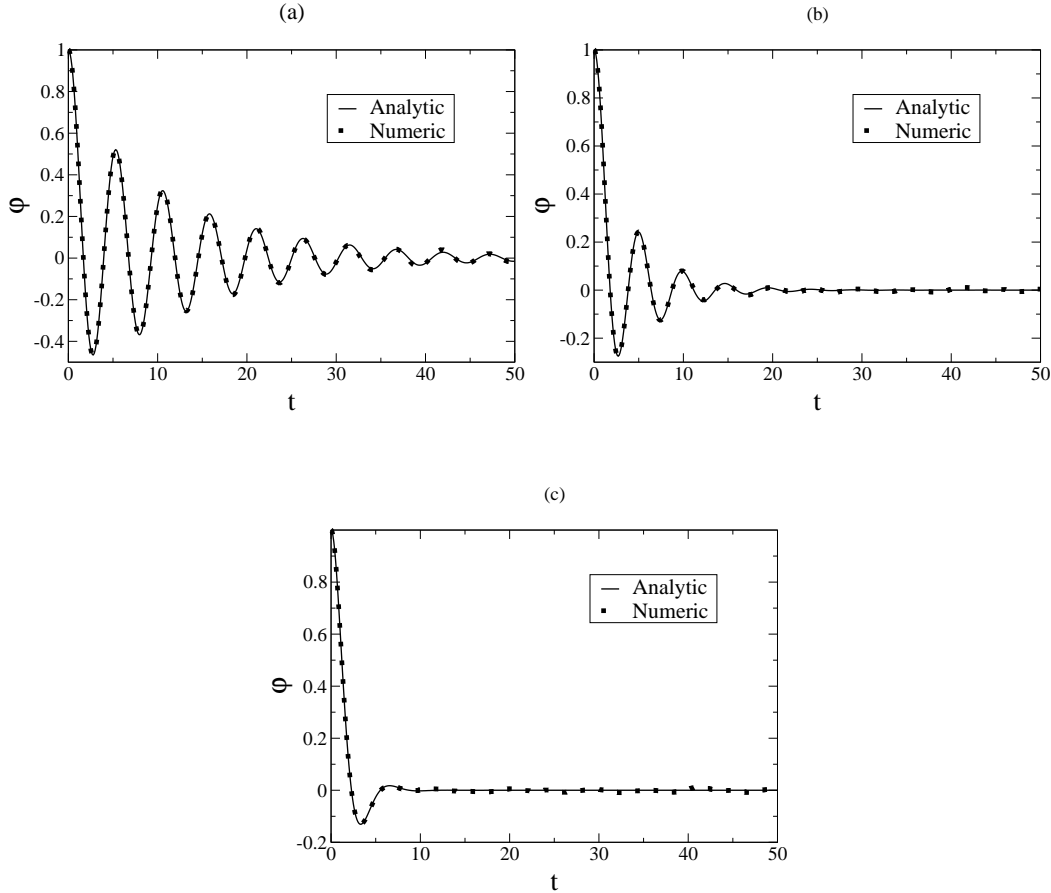

  \centerline{
    \psfig{file=fig1a.eps,scale=0.275,angle=0}
    \psfig{file=fig1b.eps,scale=0.275,angle=0}}
  \vspace{0.95 cm} \centerline{
    \psfig{file=fig1c.eps,scale=0.275,angle=0}}
  \caption{\sf Time evolution for $\varphi(t)$ in the OU case: (a) for
$\gamma=0.5$, (b) for $\gamma=1.0$ and (c) for $\gamma=5.0$. The
    other parameters are taken as $m=1.0$, $Q=1.0$ and $T=1.0$.}
  \label{fig1}
\end{figure}

\begin{figure}[htb]
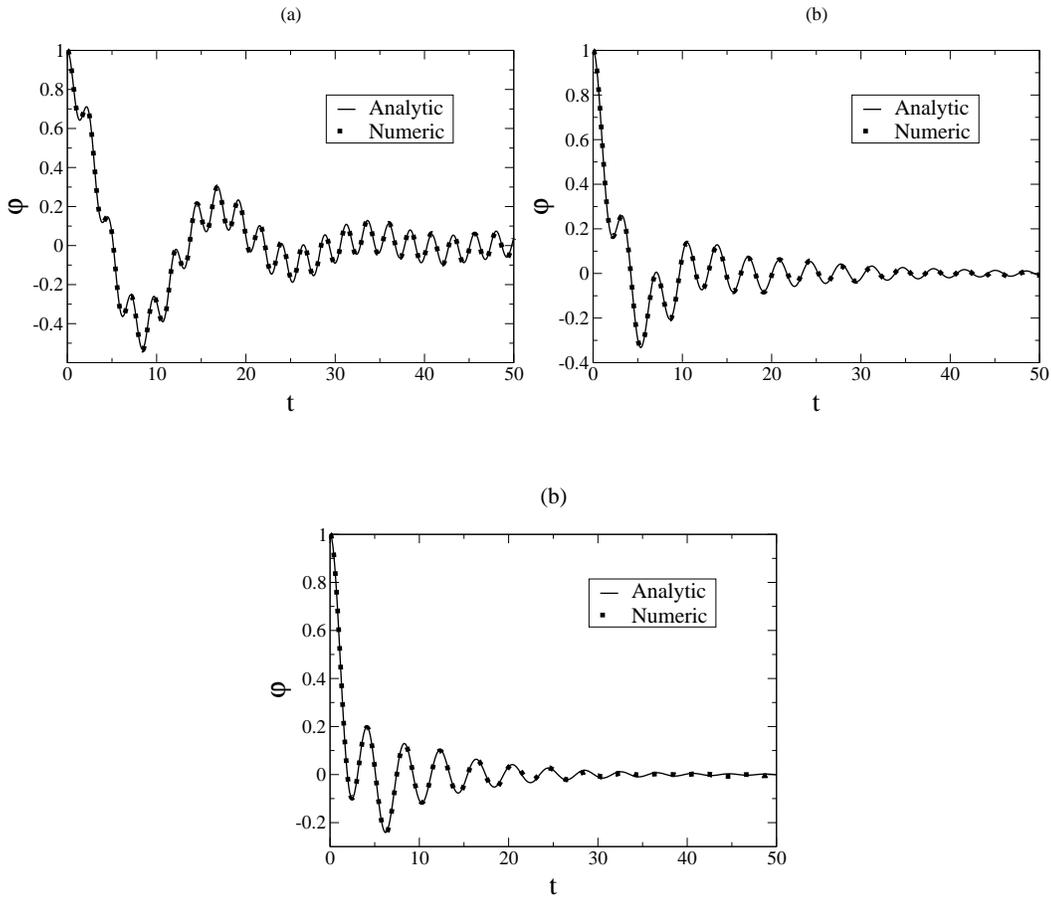

  \centerline{
    \psfig{file=fig2a.eps,scale=0.275,angle=0}
    \psfig{file=fig2b.eps,scale=0.275,angle=0}}
  \vspace{0.95 cm} \centerline{
    \psfig{file=fig2c.eps,scale=0.275,angle=0}}
  \caption{\sf Time evolution for $\varphi(t)$ in the EDH case: (a) for
$\gamma=0.1$, (b) for $\gamma=0.3$ and (c) for $\gamma=0.5$. The
    other parameters are taken as $\Omega_0=1.0$, $m=1.0$, $Q=1.0$ and $T=1.0$.}
  \label{fig2}
\end{figure}

In {}Fig.~\ref{fig1} we plot side by side our results for $\varphi(t)$
obtained from the analytical expression Eq.~(\ref{varphi}) and those obtained
numerically by solving the system of first-order differential equations,
Eq.~(\ref{setOU}), for the OU case. In {}Fig.~\ref{fig2} the same is done for
the
system Eq.~(\ref{setEDH}) for the EDH case, including the analytical solution
for this same case.  The system of differential equations (\ref{setOU}) and
(\ref{setEDH}) are solved by a standard fourth-order Runge-Kutta algorithm with
time stepsize of $\Delta t=0.01$. The number of realizations over the noise
used in both OU and EDH cases was $300,000$. In all our simulations we have
used the initial conditions $\phi(0)=1$ and $\dot\phi(0)=0$.

\begin{figure}[htb]
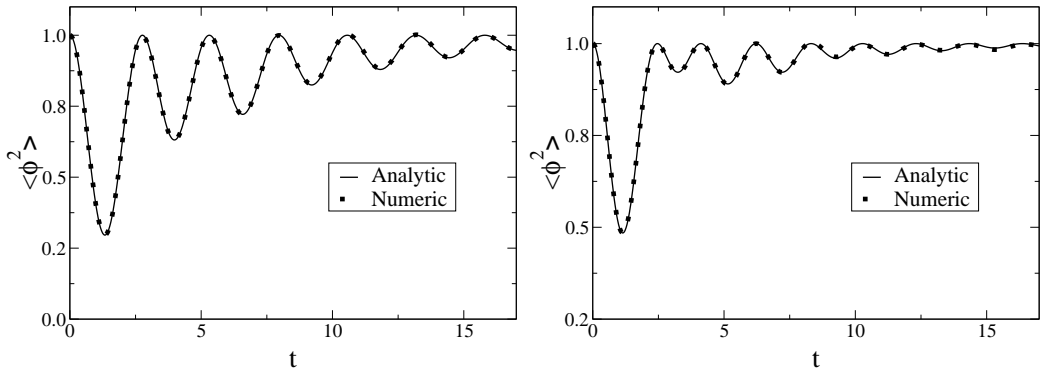

  \centerline{
    \psfig{file=fig3a.eps,scale=0.275,angle=0}
    \psfig{file=fig3b.eps,scale=0.275,angle=0}}
  \caption{\sf The time evolution for $\langle\phi^2(t) \rangle$
    in the OU case (left panel) and EDH case (right panel).  The parameters
    used are: $\gamma=0.5$, $\Omega_0=1.0$, $m=1.0$, $Q=1.0$ and $T=1.0$.}
  \label{fig3}
\end{figure}

{}From both {}Figs. \ref{fig1} and \ref{fig2} we see an excellent agreement
between the analytical and numerical results obtained from the standard
Runge-Kutta code.  {}For completeness it is also useful to compare the results
for $\langle \phi^2 \rangle$, defined analytically by Eq.~(\ref{avphi2}), with
the numerical results for this same quantity. This is done in {}Fig.~\ref{fig3}
for the OU (left panel) and EDH (right panel) cases, respectively.

{}From {}Fig.~\ref{fig3} we again see an excellent agreement between the
results obtained for $\langle \phi^2 \rangle$ analytically and numerically.
How good is this agreement between analytical and numerical results in both
cases can also be better assessed by defining the difference between them,
i.e.,

\begin{eqnarray}
\Delta \phi  &=& \varphi_{\rm analytic} - \varphi_{\rm numeric}\;
\nonumber \\
\Delta\phi^2  &=& \langle \phi^2 \rangle_{\rm analytic} -
\langle \phi^2 \rangle_{\rm numeric}\;.
\label{Deltas}
\end{eqnarray}

The results for the differences $\Delta \phi$ and $\Delta \phi^2$ are shown in
{}Figs. \ref{fig4} and \ref{fig5}, respectively, for the OU and EDH cases.

\begin{figure}[htb]
\vspace{0.5cm}
  \centerline{
    \psfig{file=fig4a.eps,scale=0.275,angle=0}
    \psfig{file=fig4b.eps,scale=0.275,angle=0}}
  \caption{\sf The difference $\Delta \phi$
    in the OU case (left panel) and EDH case (right panel).  The parameters
    used are: $\gamma=0.5$, $\Omega_0=1.0$, $m=1.0$, $Q=1.0$ and $T=1.0$.}
  \label{fig4}
\end{figure}

\begin{figure}[htb]
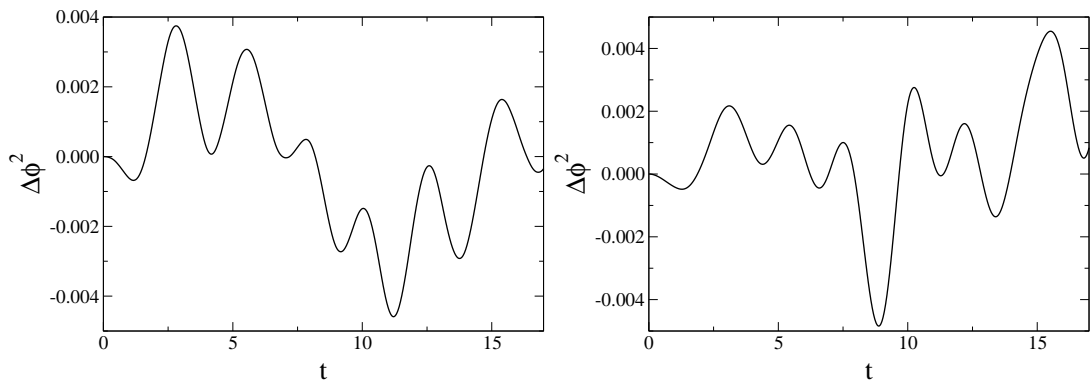

  \centerline{
    \psfig{file=fig5a.eps,scale=0.275,angle=0}
    \psfig{file=fig5b.eps,scale=0.275,angle=0}}
  \caption{\sf The difference $\Delta \phi^2$
    in the OU case (left panel) and EDH case (right panel).  The parameters
    used are: $\gamma=0.5$, $\Omega_0=1.0$, $m=1.0$, $Q=1.0$ and $T=1.0$.}
  \label{fig5}
\end{figure}

We note from the results shown in {}Figs. \ref{fig4} and \ref{fig5} that the
differences are allways smaller than about $10^{-2}$ and oscillates around zero
in a noisy way. In fact we have checked that most of this difference is purely
due to noise and can be decreased by increasing the number of realizations
over the noise. This certifies that the solution from the standard fourth-order
Runge-Kutta algorithm for the system of differential equations (\ref{setOU})
and (\ref{setEDH}) is reproducing quite well the analytical results, despite
the initial non-deterministic character of the GLE. The agreement is seen both
at short times, where the memory effects dominate,
but also at long times, where it becomes sub-dominant and where the local
approximation with dissipation Eq.~(\ref{eta}) can better represent the
dynamics \cite{bjp,markovian}.

We think that the overall error observed between the analytical
and numerical results can probably be made even smaller with an
improved code, like for example by using a stochastic Runge-Kutta
one [8]. However, we were not able to fully verify it for the particular
cases of stochastic differential equations studied here, since
its use would mean solving all equations in (19) and (23), or (26)
and (27), the same form, which seems not appropriate, since they
are not all stochastic. We hope to better discriminate this problem
in a future work, where a variation of the Runge-Kutta code is in
test to be used in situations like these.

\section{Conclusions}

In this work we have studied the reliability of using standard
numerical codes to solve generalized Langevin equations. In this
study we have used a standard fourth-order Runge-Kutta routine
to solve for the generated system of local first-order differential
equations. We have shown that the solution for the linear equation
of motion obtained from the use of a Laplace transform, when
contrasted with the numerical solution are in very good agreement
and the use of these standard numerical methods can lead to a reliable
description of the stochastic dynamics, independent of the
form of the prescription used to transform the original generalized
Langevin equation in a system of local differential equations.

We have studied the two most used cases of dissipation/noise
kernels, the OU and EDH cases. We have observed that the results
between the analytical and numerical ones agree with each other
with very good numerical precision. We expect that the numerical
precision can be made even better by using variations of a stochastic
Rung-Kutta code, appropriately tailored to deal with systems
of local differential equations like the ones we here have studied.
Work in this direction is in progress and we hope to report on the
results in a future publication.

Though we have studied in this work only the OU and EDH
non-Markovian cases, our results are expected to be of importance
for the practical study of other different generalized Langevin
equations through numerical methods, which is the case in most
situations, like when including nonlinear effects in the equations.
In those cases, once a proper prescription is used to transform
these equations, our results show that standard numerical methods
for solving differential equations can be applied to obtain reliable
results for the dynamics at both short and long times.

\acknowledgments 
The authors would like to thank Funda\c{c}\~ao de Amparo \`a
Pesquisa do Estado do Rio de Janeiro (FAPERJ) and Conselho Nacional de
Ci\^encia e Tecnologia (CNPq) for financial support.

\end{document}